\documentstyle[preprint,pre,aps,eqsecnum,epsf]{revtex}

\begin{document}
\draft

% \twocolumn[\hsize\textwidth\columnwidth\hsize\csname @twocolumnfalse\endcsname

\title{Microcanonical Lattice Gas Model for Nuclear Disassembly}

\author{C. B. Das$^1$, S. Das Gupta$^1$ and S. K. Samaddar$^2$}
\address{
$^1$Physics Department, McGill University,
3600 University St., Montr{\'e}al, Qu{\'e}bec \\ Canada H3A 2T8\\
$^2$Saha Institute of Nuclear Physics, 1/AF Bidhannagar, Calcutta 700064,
India\\ }

\date{ \today } 

\maketitle

\begin{abstract}
Microcanonical calculations are no more difficult to implement than 
canonical calculations in the Lattice Gas Model.
We report calculations for a few observables where we compare microcanonical 
model results with canonical model results.
\end{abstract}

% Maximum of 600 characters allowed for PRL

% pacs 64.60.My = metastable states}
\section{INTRODUCTION}
The Lattice Gas Model (LGM) is frequently used to compute observables
in heavy ion collisions.  The applications are numerous; see
\cite {Sam,Cho} for references.  The calculations are normally
done for a fixed temperature so that we will also call the
the usual model canonical lattice gas model CLGM.
The temperature is fixed so there will be fluctuations in energy.  It
is often argued that it might be more appropriate to keep energy
fixed.  In the following we set up a scheme for doing calculations
with fixed energy.  We will call this MLGM.

\section{Calculational Procedure}
We assume the reader is familiar with the usual LGM model for nuclear
disassembly.

In CLGM, $N$ neutrons and $Z$ protons are put in $N_s$ lattice sites
using a Metropolis algorithm.  Because of bonds between nearest neighbours
($\epsilon_{np}=-5.33$MeV, $\epsilon_{nn}=\epsilon_{pp}=0$, \cite {Sam})
and coulomb interaction between protons, there is a potential energy
which we denote by $E_{pot}$.  In Metropolis method, switch is attempted
between occupied sites and unoccupied sites and also between occupied
neutrons and protons.  If, in the switch, the energy goes down the move
is accepted.  If the energy goes up, the move is accepted but only with a 
probability $\exp (-\Delta E/T)$.  After many such switches an event
is chosen.  Once an event is picked, momenta are assigned from
Monte-Carlo sampling of a Maxwell-Boltzmann distribution at temperature
$T$.  We denote the kinetic energy by $E_{kin}$.  The total energy of
the system is $E_{tot}=E_{pot}+E_{kin}$ which will fluctuate from
event to event.

The change to MLGM can be made in the following way.  We 
start from a given configuration, hence a given $E_{pot}$.  The total
energy which will be kept fixed is $E_{tot}$, so the kinetic energy
is $E_{kin}=E_{tot}-E_{pot}$.  The available phase space of $A=N+Z$
nucleons having this kinetic energy is known analytically:
\begin{eqnarray}
\int\delta (E_{kin}-\sum_1^A p_i^2/2m)\Pi d^3p_i=\frac{2(\sqrt{\pi})^{3A}}  
{\Gamma(3A/2)}(2mE_{kin})^{3A/2-1/2}
\end{eqnarray}
We will call the value of the integral $\Omega_1(E_{kin})$.  
We now attempt a
switch in the configuration space.  The potential energy will change to
$E_{pot}'$.  To conserve energy the kinetic energy of the system should
be fixed at $E_{tot}-E_{pot}'=E_{kin}'$.  
If $\Omega_1(E_{kin}')/\Omega_1(E_{kin})>1$,
the move is accepted.  If it is less, the probablity is given by the
ratio.  Since all configurations have identical weights, 
this satisfies the principle of detailed balance.  After many such switches
an event is accepted.  We now have to assign momenta to the nucleons
so that the total kinetic energy of the $A$ nucleons add up to correct
kinetic energy which we denote by $\tilde E_{kin}$.  The correct way
to do this such that the sole criterion is phase space is the following.
Choose a sphere of radius $P$.  Do a Monte-Carlo sampling on $A$ particles
for uniform distribution in this sphere.  This means fixing $p,\theta_p,
\phi_p$ for each particle from $p=P(x_1)^{1/3}$, cos$\theta_p=1-2x_2$
and $\phi_p=2\pi x_3$ where $x_1,x_2$ and $x_3$ are random numbers.
Finally normalise $P$ so that the total energy equals
$\tilde E_{kin}$.  We are now ready to calculate all relevant quantities
including cluster decomposition.

In microcanonical simulation observables can be calculated without
having to invoke a temperature.  But it is useful to extract a 
temperature in the model.  The ``temperature'' will 
vary from event to event.  The event temperature is taken from  
$\frac{3}{2}(A-1)T\approx \frac{3}{2}AT=E_{kin}$.  The ensemble
average gives the average temperature for the given microcanonical 
total energy.  This is obviously attractive from an
experimental point of view but we can also justify it from more basic
principles.  Let us write
\begin{eqnarray}
\Omega_T(E_{tot})=\sum \Omega_1(E_{kin})\Omega_2(E_{tot}-E_{kin})
\end{eqnarray}
Although we are talking of one system only which has both kinetic and
potential energy, formally, the right hand side is the same as two
systems in ``thermal'' contact whose total energy is fixed but each 
one's individual energy can vary.  This is very standard statistical
mechanics \cite{Reif}.  For large systems, the sum is dominated by
the maximum in the product $\Omega_1(E_{kin})\Omega_2(E_{tot}-E_{kin})$
which is obtained when $\frac{\partial ln\Omega_1(E_{kin})}{\partial
E_{kin}}=\frac{\partial ln\Omega_2(E_{pot})}{\partial E_{pot}}$ which then
defines the inverse of the average temperature.  This of course leads 
again to the same identification as above and is consistent with eq.(2.1).  
The sharpness of the maximum in the 
sum depends on the size of the system.  For a small system we will
then expect the fluctuation in temperature $<T^2>-<T>^2$ to be larger
for the same value of $<T>$.

Finally although the starting point of Metropolis algorithm can be fairly
arbitrary, it is helpful computationally to start from close to 
equilibrium.  The method of generatiing the starting point followed
the scheme given
in \cite {Sam2} which was modified to take into account that there
are two kinds of bonds and the coulomb force between protons.    

Below we consider a few applications.  Many more can be made.

\section{The Caloric Curve}
As the first application, we show in Fig.1, the caloric curve, $E^*/A$ against
temperature $T$, for $^{84}$Kr (an intermediate mass disintegrating system)
and for $^{197}$Au (a heavy system).  For the microcanonical model, 
$<T>$ is used for $T$ and for the canonical one, $<E^*/A>$ is used 
for $E^*/A$. Even though
the fluctuation $<T^2>-<T>^2$ is more than four times in the case of
Kr (6.59 MeV$^2$ at $<T>=5.0$ MeV as compared to 1.1 MeV$^2$
for Au at the same average temperature) there is almost 
no difference in the $E^*/A$ vs. $T$ curves.  We should point out, 
all calculations use two kinds of bonds and include the coulomb interaction.
For $^{84}$Kr we use a cubic lattice of size $6^3$; for $^{197}$Au, we 
used $8^3$.

\section{IMF emission probability} 
One might expect that while average quantities will be nearly the same
in both the models, there would be larger differences in fluctuations
of observables.
With this in view, we have investigated Intermediate Mass Fragment ($IMF,
Z$ between 3 and 20) emissions.  We show in Fig. 2, the 
plots of $\sigma^2\equiv <n_{IMF}^2>-<n_{IMF}>^2$ (upper panel) and 
$\sigma^2/<n_{IMF}>$ (lower panel) at different temperatures.
Not a great deal of difference is found between
CLGM and MLGM calculations for intermediate or heavy masses.  
Since the fluctuation in temperature for the intermediate case
is much larger (more than a factor of four) this calls
for an explanation.  We think this is the reason.  Referring
to eq. (2.2) we see that when the temperaure is low, $E_{pot}$
is high.  That means in the particular configuration the
number of attractive bonds is less.  However, the probability
of an attractive bond being able to bind two nucleons
is much higher since the temperature is low (the energy of
relative motion $p_r^2/2\mu$ has a lesser chance of exceeding
the bond energy $-\epsilon_{np}$, see \cite {Sam}).  Similar
arguments hold when the temperature is high.  This means
$E_{pot}$ is low, so that there are many more attractive
bonds.  However, higher temperature will be able to break
these bonds more easily.  The two effects seem to cancel
each other quite efficiently.

In Fig. 3
we compare $P(n_{IMF})$, the probability of emitting $n$ IMF's, in the
two models for the case of $^{197}$Au for four different  temperatures.  
Again, the results are quite close.

We verified that in Fig.3, for temperatures 4.5 MeV and 5 MeV
both the microcanonical and canonical 
probabilities are fitted quite well by a binomial distribution:
$P(n)=\frac{m!}{n!(m-n)!}p^n(1-p)^{m-n}$ where $m$ and $p$ are
obtained from $<n>=mp$ and $\sigma^2/<n>=1-p$.  This
is a topic that has been discussed thoroughly in recent times 
\cite {Moretto97}.  This parametrisation does not work for temperatures 
3.5 MeV and 4 MeV where $\frac{\sigma^2}{<n>}>1$.  Here our calculated
points are fitted quite well by a negative binomial distribution:
$P(n)=\frac{\Gamma(N+n)}{\Gamma(N)n!}p^n(1-p)^N$.  The increase
of $\sigma^2$ at 4 MeV is not unnatural in LGM.  This is happening
because at this temperature the system is crossing the co-existence
curve \cite {Pan98} with the accompaniment of a maximum in the
fluctuation.

\section{Summary}
To summarise, microcanonical calculations in LGM are no harder to
do than canonical calculations.  The lowest mass number that we
used was 84.  Down to this size at least there are no serious
departures from canonical results.

\section{Acknowledgments}
This work was supported in part by the Natural Sciences and Engineering 
Council of Canada and by {\it le Fonds pour la Formation de chercheurs
et l'Aide \`a la Recherche du Qu\'ebec}.

\begin{figure}
\caption{Caloric curves for $^{84}Kr$ and $^{197}Au$ systems in the two
models.}
\end{figure}

\begin{figure}
\caption{${\sigma}^2\equiv <n_{IMF}^2>-<n_{IMF}>^2$ (upper panel) 
and ${\sigma}^2 / <n_{IMF}>$ (lower panel)
at different temperatures, as obtained in the two models. 
The left panels are for $^{84}Kr$ and the right ones are for $^{197}Au$.}
\end{figure}

\begin{figure}
\caption{$P(n_{IMF})$ distributions for $^{197}Au$ at $T = 3.5, 4.0,
4.5$ and $5.0 MeV$ in the two models.}
\end{figure}


\begin{references}
\bibitem{Sam} S. K. Samaddar and S. Das Gupta, Phys. Rev. {\bf C61}, 34610
(2000)

\bibitem{Cho} Ph. Chomaz and F. Gulminelli, Phys. Lett. {\bf B447}, 221
(1999)

\bibitem{Sam2} S. K. Samaddar et. al., Phys. Lett {\bf B459},8 (1999)
 
\bibitem {Reif} F. Reif, {\it Fundamentals of statistical and thermal physics}
(McGraw-Hill, New York, 1965) chap. 3

\bibitem{Moretto97} L. G. Moretto et. al. Phys. Rep. {\bf 287}, 249 (1997) 

\bibitem{Pan98} J. Pan, S. Das Gupta and M. Grant, Phys. Rev. Lett.
{\bf 80}, 1182 (1998)

\end{references}
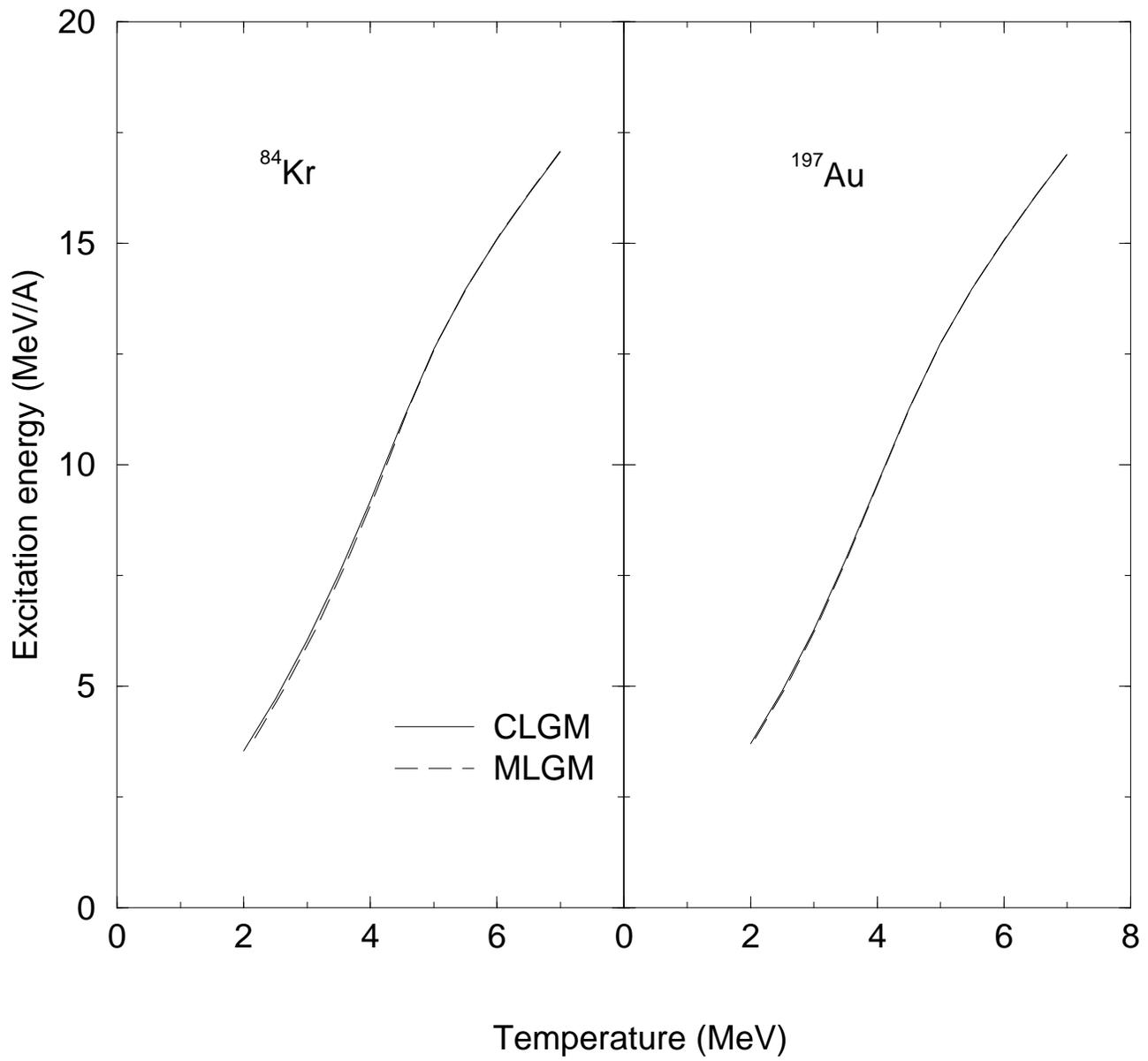
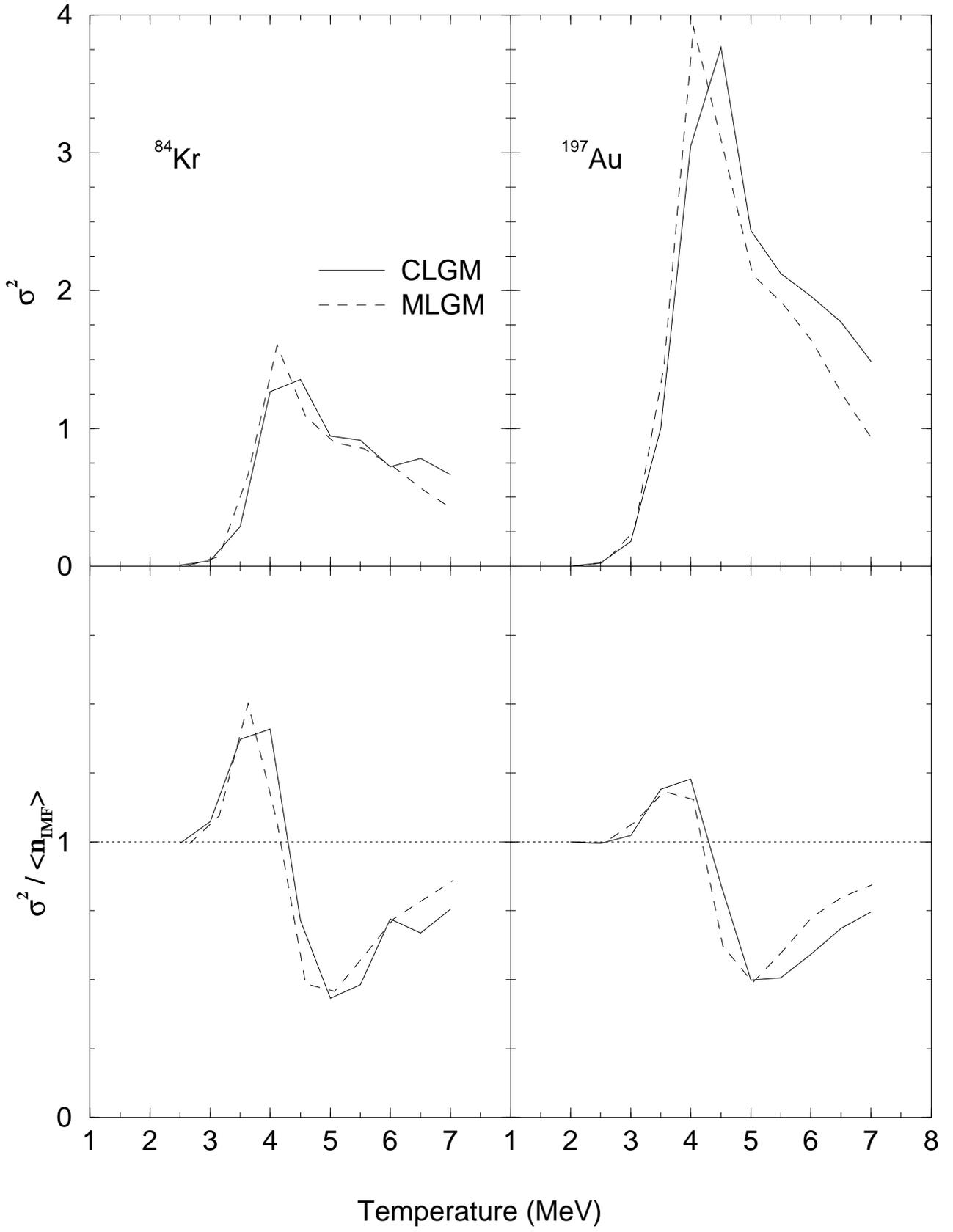
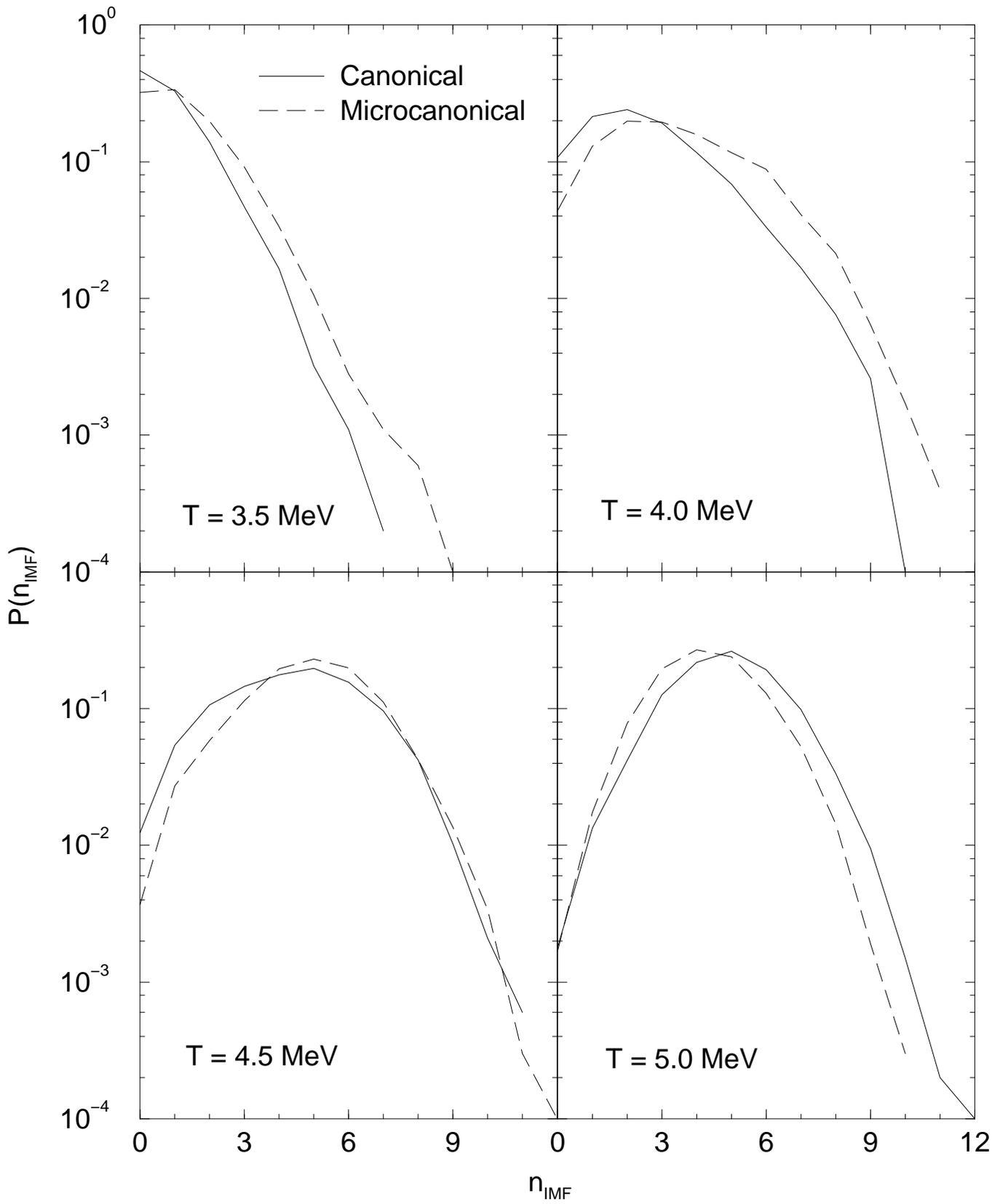
\end{document}